\documentclass[aps,prx,twocolumn,showpacs,amsmath,amssymb,superscriptaddress]{revtex4-2}
\usepackage{amsmath,amssymb,ascmac,fancybox}
\usepackage{booktabs}
\usepackage{color}
\usepackage{graphicx}
\usepackage{url}
\usepackage{siunitx}
\usepackage{bm}
\usepackage{physics}
\usepackage{tensor}
\usepackage{mathtools}
\usepackage[colorlinks=true, allcolors=blue]{hyperref}
\usepackage{txfonts}
\usepackage[compat=1.1.0]{tikz-feynhand}
\usepackage[T1]{fontenc}

\newcommand{\kB}{k_{\mathrm{B}}}
\newcommand{\ani}[1]{\hat{#1}}
\newcommand{\cre}[1]{\hat{#1}^\dagger}
\newcommand{\ene}{\varepsilon}
\newcommand{\Emat}{\mathcal{E}}
\newcommand{\Jmat}{\mathcal{J}} 
\newcommand{\JEmat}[1][]{\mathcal{J}_{\mathrm{E}{#1}}} 
\newcommand{\JQmat}[1][]{\mathcal{J}_{\mathrm{Q}{#1}}} 
\renewcommand{\vec}{\vb*}
\newcommand{\bc}{\mathcal{A}}
\newcommand{\covd}{\mathcal{D}}
\newcommand{\Vex}{V}

\newcommand{\res}{\alpha}
\newcommand{\hcvert}{\tilde{g}}

\newcommand{\eneph}{\ene_{\mathrm{ph}}}

\begin{document}
\title{Theory of shift heat current and its application to electron-phonon coupled systems}
\author{Yugo Onishi}
\affiliation{
Department of Applied Physics, University of Tokyo, 7-3-1 Hongo, Bunkyo-ku, Tokyo 113-8656
}

\author{Takahiro Morimoto}
\affiliation{
Department of Applied Physics, University of Tokyo, 7-3-1 Hongo, Bunkyo-ku, Tokyo 113-8656
}
\affiliation{JST, PRESTO, Kawaguchi, Saitama, 332-0012, Japan}

\author{Naoto Nagaosa}
\affiliation{
Center for Emergent Matter Science (CEMS), RIKEN, Wako 351-0198, Japan
}
\affiliation{
Department of Applied Physics, University of Tokyo, 7-3-1 Hongo, Bunkyo-ku, Tokyo 113-8656
}

\date{\today}

\begin{abstract}
We propose a heat current analog of the shift current, ``shift heat current''. We study nonlinear heat current responses to an applied ac electric field by a diagrammatic method and derive a microscopic expression for the second order dc heat current response. As a result, we find that the shift heat current is related to the shift vector, a geometric quantity that also appears in the expression for the shift current. The shift heat current directly depends on and can be controlled through the chemical potential. In addition, we apply the diagrammatic method to electron-phonon coupled systems, and we find that even if only the phonons are excited by an external field, the amplitude of the shift heat current is determined by the energy scale of electrons, not of phonons.
\end{abstract}

\date{\today}

\maketitle

\section{Introduction}
\begin{figure*}
    \centering
    \includegraphics[width=1.8\columnwidth]{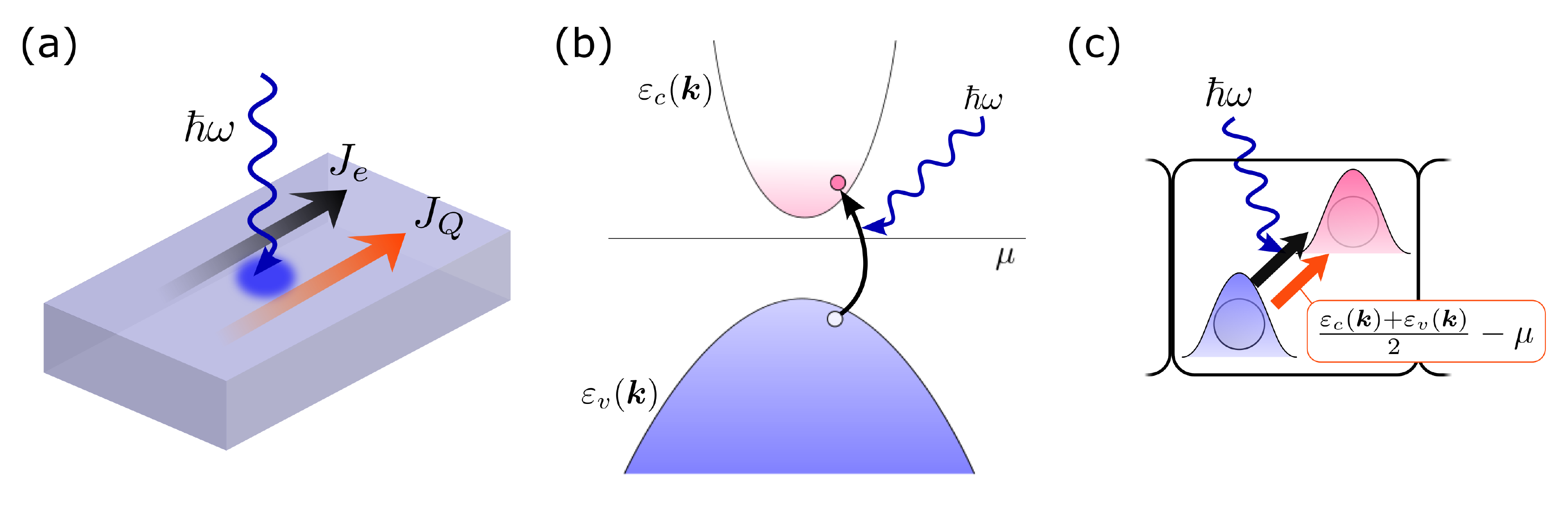}
    \caption{Schematic pictures of shift current and shift heat current. (a) Illuminating light of frequency $\omega$ induces dc electric current (shift current) and dc heat current (shift heat current). (b) In the momentum space, an incident photon excites an electron from the valence band (with dispersion $\ene_v(\vec{k})$) to the conduction band (with dispersion $\ene_c(\vec{k})$). (c) In the real space, the center of a wave packet is shifted during the interband transition. This shift of the electron wave packet (indicated by the black arrow) induces an electric current, the shift current, and simultaneously induces the shift of energy of $(\ene_c+\ene_v)/2 - \mu$ (indicated by the red arrow), which results in the shift heat current.}
    \label{fig:schematic_shift_heat_current}
\end{figure*}

The nonlinear optical response is a subject of recent intensive research. For example, second order optical responses such as second harmonic generation have been attracting much attention and studied both theoretically and experimentally \cite{Belinicher1978a, Ivchenko1978, VonBaltz1981, Belinicher1982, Eaton1991, sipe2000, Young2012, Young2012a, Morimoto2016, Tan2016, Wu2017, Cook2017, Nagaosa2017, Kim2017, Nakamura2017, Ogawa2017, Parker2019, Osterhoudt2019a, Sotome2019, Ahn2020, Golub2020, Avdoshkin2020}. 
Recently, it is also pointed out that even in nonlinear regime quantum geometry and topological properties of materials are important to understand the electric/optical responses \cite{Morimoto2016, Ahn2020}. Among them, the bulk photovoltaic effect is an important issue for both applications and fundamental physics. The bulk photovoltaic effect, or sometimes called photogalvanic effect, is the generation of photocurrents that can occur in noncentrosymmetric materials, and a mechanism called shift current is proposed and well established \cite{VonBaltz1981, Young2012, Young2012a, Morimoto2016, Nagaosa2017, Ahn2020}. The shift current is characterized by a quantity called shift vector $\vec{R}$, and for two bands systems with the time reversal symmetry (TRS), the shift current $J_e^{\mathrm{(shift)}}$ induced by an electric field $E$ can be written as \cite{VonBaltz1981, Morimoto2016}
\begin{align}
	&J_e^{\mathrm{(shift)}} \propto E^2\int\dd{\vec{k}}\delta(\ene_v(\vec{k})- \ene_c(\vec{k}) + \hbar\Omega) \abs{v_{vc}(\vec{k})}^2 R(\vec{k}),
\end{align}
where $\ene_c(\vec{k})$ and $\ene_v(\vec{k})$ are the dispersion of the conduction band and the valence band, $\hbar\Omega$ is the energy of the input photon, $v_{vc}(\vec{k})$ is the matrix element of the velocity operator. Physically, the shift vector $\vec{R}$ can be interpreted as the spatial shift of an electron wave packet during the interband transition due to an excitation by light (Fig.~\ref{fig:schematic_shift_heat_current}(b,c)), and contributes to the dc electric current. The shift of wave packet can contribute other types of transport. Kim {\it et al.} pointed out that the shift can contribute to spin current and proposed a phenomenon named shift spin current~\cite{Kim2017}. In the present work, we propose another current induced by the shift: shift heat current.

Electrons carry not only charge and spin, but also heat. Therefore, it is natural to expect that there exists a heat current analog of shift current. In fact, heat transport phenomena and thermoelectric responses are closely related to electric/optical responses. It is well known that several universal relations between thermal responses and electric responses, such as the Wiedemann-Franz law and the Mott relation, hold \cite{Behnia2015}. These universal relations can be derived theoretically by the semiclassical Boltzmann theory \cite{Behnia2015}, or quantum linear response theory \cite{Kubo1957, Kubo1957a, Luttinger1964, Ogata2019}. Within the linear response regime, the Mott relation or the Sommerfeld-Bethe relation \cite{Ogata2019} indicates that the amplitude of the electric current $J_e$ and the heat current $J_Q$ when an electric field is applied can be roughly related to each other as 
\begin{align}
	J_Q \sim \frac{t}{e} J_e, \label{eq:J_JQ_relation}
\end{align}
where $e$ is the charge of an electron and $t$ is the characteristic energy scale an electron carries as heat. If we assume that the relation between $J_e$ and $J_Q$ in Eq.~\eqref{eq:J_JQ_relation} holds even for nonlinear responses, there should be nonlinear heat responses corresponding to the shift current, namely, shift heat current. One can also expect that the shift heat current are related to the geometric property of the material as the shift current does. 

Although the electric current responses and heat current responses are closely related, the theoretical investigation on heat transport and thermoelectric effects has not progressed compared with that of the electric conductivity or the optical responses. In particular, thermoelectric responses of macroscopic systems in nonlinear regime are rarely discussed in contrast to aforementioned optical responses. 
In previous studies, nonlinear thermoelectric effects were studied mainly for mesoscopic systems \cite{Sanchez2016b, Sanchez2013, Leijnse2010, Karbaschi2016, Thierschmann2015, Lopez2013}, where large temperature bias can be relatively easily applied. Although there are several studies on nonlinear thermoelectric responses to dc external fields in macroscopic systems \cite{Bhalla2021, Osada2021, Gao2018, Hwang2014}, heat current responses to ac external fields including the shift current-like responses have not been studied so far. 

In this work, we study the dc heat current responses to an ac electric field. We extend the diagrammatic method to calculate nonlinear responses proposed in Ref.~\cite{Parker2019} to heat current responses, and find that the relation in Eq.~\eqref{eq:J_JQ_relation} also holds for the shift current. In particular, for two-band systems with TRS, the shift heat current response $J_Q^{\mathrm{(shift)}}$ can be written as 
\begin{align}
	J_Q^{\mathrm{(shift)}} \propto & E^2\int\dd{\vec{k}}\delta(\ene_v(\vec{k})- \ene_c(\vec{k}) + \hbar\Omega) \a	bs{v_{vc}(\vec{k})}^2 R(\vec{k}) \nonumber \\
	&\times\qty(\frac{\ene_c(\vec{k}) + \ene_v(\vec{k})}{2} -\mu). \label{eq:JQ_shift_prop}
\end{align}
We also find that the shift heat current directly depends on the chemical potential $\mu$ and thus we can control the shift heat current through the chemical potential.

Furthermore, we also discuss the phonon-induced shift heat current. Recently, the phonon-induced shift current is experimentally observed in BaTiO$_3$ \cite{Okamura2022}. There, although the energy of the input photon ($\sim\SI{1}{\milli\electronvolt}$) is much smaller than the band gap of BaTiO$_3$ ($\sim\SI{1}{\electronvolt}$), phonons are excited by photons and then induce a large electric current of the order of $\SI{10}{\micro\ampere}$. Although the mechanism is different from the usual shift current, we can also expect the similar relation as Eq.~\eqref{eq:J_JQ_relation}. We theoretically derive the expression for such electron-phonon coupled systems, and find again that the relation~\eqref{eq:J_JQ_relation} holds. It should be noted that, although photons excite phonons only, the energy scale $t$ in Eq.~\eqref{eq:J_JQ_relation} is still that of electrons, not of phonons.

This paper is organized as follows. In Sec.~\ref{sec:calc_method}, we describe the theoretical formalism used to calculate the shift heat current. In Sec.~\ref{sec:shift_heat_current}, we present the calculation results for the shift heat current. In addition to the most general formula of shift heat current, formulas for systems with TRS and TR symmetric two band systems are also given. In Sec.~\ref{sec:Rice-Mele}, we apply our theory to Rice-Mele model as an example. In Sec.~\ref{sec:phonon_induced_shift_heat_current}, we study the phonon-induced shift heat current. Sec.~\ref{sec:discussion} is devoted to discussions.

\section{Calculation method} \label{sec:calc_method}
In this paper, we calculate the nonlinear responses of the heat current to an electric field. To this end, we follow the formalism adopted in Ref.~\cite{Parker2019}. We first summarize our notations regarding the basics of the band theory. We introduce Bloch Hamiltonian and covariant derivative of $\vec{k}$-dependent operators, which is convenient to expand the Bloch Hamiltonian with respect to the applied external field. Then we introduce the electric current operator, the energy current operator, and the heat current operator.

\subsection{Band theory and covariant derivative}
Let us consider a tight binding model of noninteracting electron systems in a periodic potential, written as 
\begin{align}
	\hat{H}_0 &= \sum_{\vec{R},\vec{R}'}\cre{\psi}_{\vec{R}} H_{\vec{R}-\vec{R}'}\ani{\psi}_{\vec{R}'}, \label{eq:Hamiltonian}\\
	\ani{\psi}_{\vec{R}} &= \qty(\ani{\psi}_{\vec{R},1}, \ani{\psi}_{\vec{R},2}, \dots, \ani{\psi}_{\vec{R},s})^T, \\
	\cre{\psi}_{\vec{R}} &= \qty(\cre{\psi}_{\vec{R},1}, \cre{\psi}_{\vec{R},2}, \dots, \cre{\psi}_{\vec{R},s}). 
\end{align}
Here, $\vec{R}$ specifies the position of a unit cell. $\ani{\psi}_{\vec{R}, i}$ ($\cre{\psi}_{\vec{R}, i}$) is an annihilation (creation) operator, which annihilates(creates) an electron in the $i$-th orbital of the unit cell at $\vec{R}$. $H_{\vec{R}-\vec{R}'}$ is an $s\times s$ matrix. Note that we denote the operator in the Fock space a symbol with a hat (~$\hat{}$~) while its matrix representation is denoted by a symbol without $\hat{}$ .

The Hamiltonian in the momentum space representation is 
\begin{align}
	\hat{H}_0 &= \sum_{\vec{k}} \cre{\psi}_{\vec{k}} H_0(\vec{k}) \ani{\psi}_{\vec{k}}, \\
	(H_0(\vec{k}))_{ij} &= \sum_{\vec{R}} e^{-i\vec{k}\vdot(\vec{R}+\vec{r}_i-\vec{r}_j)} (H_{\vec{R}})_{ij},
\end{align}
where $\ani{{\psi}}_{\vec{k},i} = \dfrac{1}{\sqrt{N}}\sum_{\vec{R}} e^{-i\vec{k}\vdot(\vec{R}+\vec{r}_i)} \ani{\psi}_{\vec{R},i}$, $N$ is the total number of unit cells and $\vec{R}+\vec{r}_i$ is the position of the $i$-th orbital in the unit cell at $\vec{R}$.

By diagonalizing the Hamiltonian, one obtain the representation in the energy eigenstate basis:
\begin{align}
	\Emat_{\vec{k}} &= U_{\vec{k}}^\dagger H_0(\vec{k}) U_{\vec{k}},
\end{align}
where $U_{\vec{k}}$ is a unitary matrix and $\Emat_{\vec{k}}$ is a diagonal matrix and its element $(\Emat_{\vec{k}})_{ab} = \delta_{ab} \ene_{\vec{k},a}$ gives the dispersion of the $a$-th band.	Then the Hamiltonian can be written as 
\begin{align}
	\hat{H}_0 &= \sum_{\vec{k}} \cre{c}_{\vec{k}} \Emat_{\vec{k}} \ani{c}_{\vec{k}},
\end{align}
where $\ani{\psi}_{\vec{k}} = U_{\vec{k}}\ani{c}_{\vec{k}}$.

For convenience, we introduce a covariant derivative. Consider an operator of the following form:
\begin{align}
	\hat{O} &= \sum_{k} \cre{\psi}_{\vec{k}} O(\vec{k}) \ani{\psi}_{\vec{k}} = \sum_{k} \cre{c}_{\vec{k}} \mathcal{O}_{\vec{k}} \ani{c}_{\vec{k}}, \\
	\mathcal{O}_{\vec{k}} &= U_{\vec{k}}^\dagger O(\vec{k}) U_{\vec{k}}.
\end{align}
The covariant derivative appears when one considers the $k$-derivative of the operator, namely,  
\begin{align}
	\vec{D}[\hat{O}] &:= \sum_{k} \cre{\psi}_{\vec{k}} \nabla_{\vec{k}}O(\vec{k}) \ani{\psi}_{\vec{k}}.
\end{align}
Substituting $\ani{\psi}_{\vec{k}} = U_{\vec{k}}\ani{c}_{\vec{k}}$ leads to the expression in the energy eigen basis,
\begin{align}
	\vec{D}[\hat{O}] &= \sum_{k} \cre{c}_{\vec{k}} \vec{\covd}[\mathcal{O}_{\vec{k}}] \ani{c}_{\vec{k}} ,\\
	\vec{\covd}[O_{\vec{k}}] &:= \nabla_{\vec{k}}\mathcal{O}_{\vec{k}} - i\commutator{\vec{\bc}_{\vec{k}}}{\mathcal{O}_{\vec{k}}},
\end{align}
where $\vec{\bc}_{\vec{k}}$ is the interband Berry connection defined as $\vec{\bc}_{\vec{k}} = iU_{\vec{k}}^\dagger \nabla_{\vec{k}} U_{\vec{k}}$. Here, $\vec{\covd}$ is the covariant derivative. We note that the Leibniz rule holds for the covariant derivative:
\begin{align}
	\covd^{\alpha}[\mathcal{O}_1\mathcal{O}_2] = (\covd^{\alpha}[\mathcal{O}_1])\mathcal{O}_2 + \mathcal{O}_1(\covd^{\alpha}[\mathcal{O}_2]). \label{eq:Leibniz_rule}
\end{align}

\subsection{Electromagnetic interaction}
Next, let us consider the Hamiltonian describing the electromagnetic interaction, and expand it with respect to the applied electric field.

To express a spatially uniform external electric field $\vec{E}(t)$, we use the velocity gauge. In this gauge, the electromagnetic interaction with a spatially uniform electric field does not violate the translational symmetry and is incorporated by the minimal substitution as
\begin{align}
	\hat{H}_A(t) &= \sum_{\vec{k}} \cre{\psi}_{\vec{k}} H_A(\vec{k}) \ani{\psi}_{\vec{k}},\\
	H_A(\vec{k}) &= H_0\qty(\vec{k}-\frac{e}{\hbar}\vec{A}(t)),
\end{align}
where $e<0$ is the charge of an electron, and $\vec{A}(t)$ is a vector potential satisfying $\partial \vec{A}(t)/\partial t = -\vec{E}(t)$.

To treat the external field perturbatively, we expand the Hamiltonian with respect to $\vec{A}(t)$ up to second order, 
\begin{align}
    H_A(\vec{k}) &= H_0(\vec{k}) 
     + \qty(-\frac{e}{\hbar})A^{\alpha}(t) \partial_{\alpha_k}H_0(\vec{k}) \nonumber\\
     &\qquad + \frac{1}{2}\qty(-\frac{e}{\hbar})^2A^{\alpha}(t)A^{\beta}(t)\partial_{\alpha}\partial_{\beta}H_0(\vec{k}).
\end{align} 
Here $\partial_{\alpha}=\partial/\partial_{k_{\alpha}}$ and $\alpha_k$ is a spatial index. (In this paper, greek indices $\mu,\alpha,\beta,\dots$ always represent the spatial indices with an implicit summation henceforth).
This expression is in the basis $\ani{\psi}_{\vec{k}}$. For analytic calculation, it is convenient to move to the energy eigenstate basis, $\ani{c}_{\vec{k}}$. As already discussed, the $k$-derivative in the basis $\ani{\psi}_{\vec{k}}$ becomes the covariant derivative $\vec{\covd}$ in the basis $\ani{c}_{\vec{k}}$. Therefore, up to $\order{A^2}$, the Hamiltonian reads
\begin{align}
	\hat{H}_A(t) 
	&= \hat{H}_0 + \hat{\Vex}, \\
	\hat{\Vex} &= \sum_{\vec{k}} \cre{c}_{\vec{k}} (\Vex_{1,\vec{k}}+\Vex_{2,\vec{k}}) \ani{c}_{\vec{k}}, 
\end{align}
with
\begin{align}
    \Vex_{1,\vec{k}}&= \qty(-\frac{e}{\hbar})A^{\alpha}(t) h^{\alpha}_{\vec{k}} 
    = \int\frac{\dd{\omega}}{2\pi} \frac{ie}{\hbar\omega_k}E^{\alpha}(\omega)e^{-i\omega t} h^{\alpha}_{\vec{k}} \\ 
    \Vex_{2,\vec{k}}&= \frac{1}{2}\qty(-\frac{e}{\hbar})^2A^{\alpha}(t)A^{\beta}(t) h^{\alpha\beta}_{\vec{k}} \nonumber \\
    &= \frac{1}{2}\int\frac{\dd{\omega_1}}{2\pi}\frac{ie}{\hbar\omega_1}E^{\alpha}(\omega_1)e^{-i\omega_1 t}\int\frac{\dd{\omega_2}}{2\pi}\frac{ie}{\hbar\omega_2}E^{\beta}(\omega_2)e^{-i\omega_2 t} h^{\alpha\beta}_{\vec{k}}
\end{align}
where $h^{\alpha_1\dots\alpha_n}$ is defined as 
\begin{align}
	h^{\alpha_1\dots\alpha_n} &= \covd^{\alpha_1}\dots\covd^{\alpha_n}[\Emat_{\vec{k}}].
\end{align}
Although $h^{\alpha_1\dots\alpha_n}$ depends on $\vec{k}$, we omit $\vec{k}$ from its notation. (We often omit $\vec{k}$-dependencies in other quantities as well in the following.) $E^{\alpha}(\omega)$ is the Fourier transform of the electric field and is related to the Fourier transform of the vector potential $A^{\alpha}(\omega)$ as 
\begin{align}
	E^{\alpha}(\omega) &= \int\dd{t} e^{i\omega t} E^{\alpha}(t) = i\omega A^{\alpha}(\omega).
\end{align}

\subsection{Particle current operator, energy current operator, and heat current operator}
Here we will introduce current operators used in this work. The current operators defined below change their form in the presence of an external electric field, and thus we expand them with respect to the external field, as we did for the Hamiltonian in the previous section.

It is well known that the total particle current operator $\hat{J}$ is given by the $k$-derivative of the Hamiltonian, i.e.,
\begin{align}
	\hat{\vec{J}} &= \frac{1}{V} \sum_{\vec{k}} \cre{\psi}_{\vec{k}} \frac{1}{\hbar}\nabla_{\vec{k}}H_A(\vec{k}) \ani{\psi}_{\vec{k}} \label{eq:current_operator_def},
\end{align}
where $V$ is the volume of the system.
Note that the application of an electric field changes the form of the electric current operator in this gauge. This can be rewritten in terms of the covariant derivative and $\ani{c}_{\vec{k}}$ as 
\begin{align}
	&\hat{J}^{\mu} = \sum_{\vec{k}} \cre{c}_{\vec{k}} \Jmat^\mu(\vec{k}) \ani{c}_{\vec{k}} = \sum_{\vec{k}} \cre{c}_{\vec{k}} (\Jmat^\mu_0(\vec{k}) + \Jmat^\mu_1(\vec{k}) + \Jmat^\mu_2(\vec{k})) \ani{c}_{\vec{k}},
\end{align}
with
\begin{align}
    \Jmat^\mu_0(\vec{k}) &= \frac{1}{\hbar} h^{\mu}, \\
    \Jmat^\mu_1(\vec{k}) &= \frac{1}{\hbar} \int\frac{\dd{\omega}}{2\pi} \frac{ie}{\hbar\omega}E^{\alpha}(\omega)e^{-i\omega t}h^{\mu\alpha}, \\
    \Jmat^\mu_2(\vec{k}) 
    & = \frac{1}{2\hbar} \int\frac{\dd{\omega_1}}{2\pi} \frac{ie}{\hbar\omega_1}E^{\alpha}(\omega_1)e^{-i\omega_1 t} \nonumber \\
    & \quad \times \int\frac{\dd{\omega_2}}{2\pi} \frac{ie}{\hbar\omega_2}E^{\beta}(\omega_2)e^{-i\omega_2 t} h^{\mu\alpha\beta}.
\end{align}
(Note again that we omit $\vec{k}$ from the notation of $h^{\alpha_1\dots\alpha_n}$.)

When we ignore the interaction between electrons, the total energy current operator in longitudinal responses to an external electric field is given by
\begin{align}
	\hat{\vec{J}_E} &= \frac{1}{V} \sum_{\vec{k}} \cre{\psi}_{\vec{k}} \frac{1}{2}(H_A(\vec{k})\vec{\Jmat}(\vec{k}) + \vec{\Jmat}(\vec{k})H_A(\vec{k})) \ani{\psi}_{\vec{k}} \nonumber \\
	&= \frac{1}{V} \sum_{\vec{k}} \cre{\psi}_{\vec{k}} \frac{1}{2\hbar}\nabla_{\vec{k}}(H_A(\vec{k})^2)\ani{\psi}_{\vec{k}}. \label{eq:energy_current_operator_def}
\end{align} 
Equation~\eqref{eq:energy_current_operator_def} shows that the application of an electric field changes the form of the energy current operator in the velocity gauge. This is similar to the electric current operator, which also changes its form in the presence of a vector potential.

The energy current operator can be rewritten in terms of the covariant derivative and $\ani{c}_{\vec{k}}$ as
\begin{align}
	\hat{J}_E^{\mu}&= \frac{1}{V} \sum_{\vec{k}} \cre{c}_{\vec{k}} \JEmat^\mu \ani{c}_{\vec{k}} = \frac{1}{V} \sum_{\vec{k}} \cre{c}_{\vec{k}} (\JEmat[0]^\mu + \JEmat[1]^\mu + \JEmat[2]^\mu)\ani{c}_{\vec{k}}, \label{eq:JE_expression}
\end{align}
with
\begin{align}
	\JEmat[0]^\mu(\vec{k}) &= \frac{1}{\hbar} g^{\mu}, \\
    \JEmat[1]^\mu(\vec{k}) &= \frac{1}{\hbar} \int\frac{\dd{\omega}}{2\pi} \frac{ie}{\hbar\omega}E^{\alpha}(\omega)e^{-i\omega t}g^{\mu\alpha}, \\
    \JEmat[2]^\mu(\vec{k}) 
    & = \frac{1}{2\hbar} \int\frac{\dd{\omega_1}}{2\pi} \frac{ie}{\hbar\omega_1}E^{\alpha}(\omega_1)e^{-i\omega_1 t} \nonumber\\ 
    & \quad \times \int\frac{\dd{\omega_2}}{2\pi} \frac{ie}{\hbar\omega_2}E^{\beta}(\omega_2)e^{-i\omega_2 t} g^{\mu\alpha\beta},
\end{align}
where $g^{\mu\alpha_1\dots\alpha_n}$ is defined as 
\begin{align}
	g^{\alpha_1\dots\alpha_n} &= \frac{1}{2}\covd^{\alpha_1}\dots\covd^{\alpha_n}[\Emat_{\vec{k}}^2].
\end{align}

The heat current operator $\hat{\vec{J}}_Q$ is defined as 
\begin{align}
	\hat{\vec{J}}_Q &\equiv \hat{\vec{J}}_E - \mu \hat{\vec{J}} \nonumber\\
	&=  \frac{1}{V}\sum_{\vec{k}} \cre{\psi}_{\vec{k}} \frac{1}{\hbar}\nabla_{\vec{k}}\qty(\frac{1}{2}H_A(\vec{k})^2-\mu H_A(\vec{k}))\ani{\psi}_{\vec{k}} \nonumber\\
	&=  \frac{1}{V}\sum_{\vec{k}} \cre{\psi}_{\vec{k}} \frac{1}{2\hbar}\nabla_{\vec{k}}(H_A(\vec{k})-\mu)^2 \ani{\psi}_{\vec{k}}, \label{eq:heat_current_operator_def}
\end{align}
where $\mu$ is the chemical potential. Therefore, one obtains the heat current operator by replacing $\Emat_{\vec{k}}$ in Eq.~\eqref{eq:JE_expression} with $\Emat_{\vec{k}} - \mu$,
which yields
\begin{align}
	\hat{J}_Q^{\mu}&= \frac{1}{V} \sum_{\vec{k}} \cre{c}_{\vec{k}} \JQmat^\mu \ani{c}_{\vec{k}} = \frac{1}{V} \sum_{\vec{k}} \cre{c}_{\vec{k}} (\JQmat[0]^\mu + \JQmat[1]^\mu + \JQmat[2]^\mu)\ani{c}_{\vec{k}}, \label{eq:JQ_expression}
\end{align}
with
\begin{align}
	\JQmat[0]^\mu(\vec{k}) &= \frac{1}{\hbar} \hcvert^{\mu}, \\
    \JQmat[1]^\mu(\vec{k}) &= \frac{1}{\hbar} \int\frac{\dd{\omega}}{2\pi} \frac{ie}{\hbar\omega}E^{\alpha}(\omega)e^{-i\omega t}\hcvert^{\mu\alpha}, \\
    \JQmat[2]^\mu(\vec{k})
    & = \frac{1}{2\hbar} \int\frac{\dd{\omega_1}}{2\pi} \frac{ie}{\hbar\omega_1}E^{\alpha}(\omega_1)e^{-i\omega_1 t} \nonumber \\ &\quad \times \int\frac{\dd{\omega_2}}{2\pi} \frac{ie}{\hbar\omega_2}E^{\beta}(\omega_2)e^{-i\omega_2 t} \hcvert^{\mu\alpha\beta}, 
\end{align}
and
\begin{align}
	&\hcvert^{\alpha_1\dots \alpha_n} = \frac{1}{2}\covd^{\alpha_1}\dots\covd^{\alpha_n}[(\Emat_{\vec{k}}-\mu)^2]. \label{eq:heat_current_vertex}
\end{align}

Our goal is to calculate the expectation value $\expval{\hat{J}^\mu_Q(t)}$. The heat current responses are characterized by the tensors $\res^{\mu\alpha_1\alpha_2\dots\alpha_n}(t; t_1, t_2,\dots t_n)$ as
\begin{align}
	\expval{\hat{J}_Q^\mu(t)} &= \sum_{n=0}^\infty \frac{1}{n!} \int\qty[\prod_{k=1}^n\dd{t_k}E^{\alpha_k}(t_k)] \res^{\mu\alpha_1\alpha_2\dots\alpha_n}(t; t_1, t_2,\dots t_n).
\end{align}
By Fourier transformation, we can write
\begin{align}
	&\expval{\hat{J}_Q^\mu(\omega)} = \int\dd{t} e^{i\omega t} \expval{\hat{J}_Q^\mu(t)} \nonumber \\
	&= \sum_{n=0}^\infty \frac{1}{n!} \int \qty[\prod_{k=1}^n\frac{\dd{\omega_k}}{2\pi}E^{\alpha_k}(\omega_k)] \res^{\mu\alpha_1\alpha_2\dots\alpha_n}(\omega; \omega_1, \omega_2,\dots \omega_n), \\
	&\res^{\mu\alpha_1\alpha_2\dots\alpha_n}(\omega; \omega_1, \omega_2,\dots \omega_n) \nonumber \\
	&= \int\frac{\dd{t}}{2\pi}e^{i\omega t} \int\qty[\prod_{k=1}^n \frac{\dd{t_k}}{2\pi} e^{-i\omega_k t_k}] \res^{\mu\alpha_1\alpha_2\dots\alpha_n}(t; t_1, t_2,\dots t_n).
\end{align}
In particular, the second order response coefficient $\res^{\mu\alpha\beta}(\omega; \omega_1, \omega_2)$ can be written with functional derivatives as 
\begin{align}
	&\res^{\mu\alpha\beta}(\omega; \omega_1, \omega_2) \nonumber \\
	&= \int\frac{\dd{t}}{2\pi} \int\frac{\dd{t_1}}{2\pi}\int\frac{\dd{t_2}}{2\pi} e^{i(\omega t - \omega_1 t_1 - \omega_2 t_2)} \eval{\frac{\delta}{\delta E^{\alpha}(t_1)}\frac{\delta}{\delta E^{\beta}(t_2)}\expval{\hat{J}_Q^\mu (t)}}_{\vec{E}=0}.
\end{align}
Note that we need to take functional derivatives in the time domain because we are considering nonlinear regime as in the case of the electric current responses \cite{Parker2019}.

We notice a formal similarity between the current operator and the energy/heat current operator. The current operator (Eq.~\eqref{eq:current_operator_def}) is given by the $k$-derivative of the Hamiltonian, while the energy current operator is the $k$-derivative of the squared Hamiltonian. By replacing the Hamiltonian $H$ in the energy current operator with $H-\mu$, we can also obtain the heat current operator. Because of this similarity, we can calculate the heat current response in the same manner as the one used for the shift current. Because the rest of the formulation is almost the same as Ref.~\cite{Parker2019}, we only present the results in the next section. 

\section{Shift heat current} \label{sec:shift_heat_current}

In this section, we present results for the second order dc responses of heat current with respect to an electric field with frequency $\Omega$, which we call ``shift heat current''. We derive a general expression for the shift heat current, and then reduce the expression to cases for time reversal invariant systems, and especially, time reversal symmetric two band systems. 

\subsection{General expression for shift heat current}
We focus on the second order dc responses to an electric field with the frequency $\Omega$. We also restrict ourselves to longitudinal responses, namely, $\res^{xxx}(2i\eta;\Omega+i\eta, -\Omega+i\eta)$ where $\eta$ is an infinitesimal positive quantity. First we calculate  $\res^{xxx}$ within the imaginary time formalism, and then continue the result to the real time expressions. From now on, we set $\hbar=1$ for simplicity. 

\begin{figure}[tbp]
    \centering
    \includegraphics[width=\columnwidth]{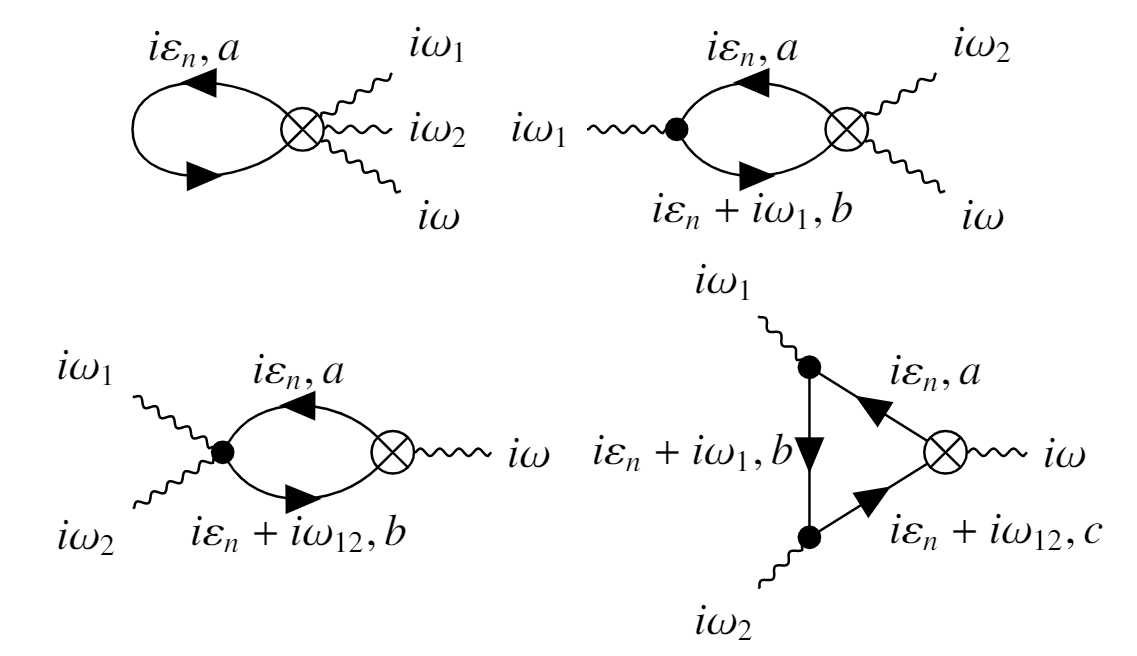}
    \caption{The Feynman diagrams which contribute to shift heat current. Solid lines with an arrow and wavy lines represent propagators of electrons and photons, respectively, and black dots with $n$ photons denote vertices for $h^{\alpha_1\dots\alpha_n}$ and cross dots with $n$ photons are the vertex for $\hcvert^{\alpha_1\dots\alpha_n}$.}
    \label{fig:diagram_shift_heat_current}
\end{figure}

The Feynman diagrams which contribute to the second order heat current responses are shown in Fig.~\ref{fig:diagram_shift_heat_current}.
After performing the Matsubara frequency summation, we obtain
\begin{widetext}
\begin{align}
	&\res^{xxx}(2i\eta; \Omega+i\eta, -\Omega+i\eta) \nonumber\\
	&= \frac{1}{V}\sum_{\vec{k}}\frac{e^2}{\Omega^2} 
	\left[\sum_{a}\hcvert_{aa}^{xxx} f_a + \sum_{a,b}\hcvert_{ab}^{xx} h_{ba}^x \qty(\frac{f_{ab}}{\ene_{ab}+\Omega+i\eta} + \frac{f_{ab}}{\ene_{ab}-\Omega+i\eta}) 
	 + \sum_{a,b}\hcvert_{ab}^{x} h_{ba}^{xx} \frac{f_{ab}}{\ene_{ab}+2i\eta} \right. \nonumber \\
	& \left. + \sum_{a,b,c} \frac{\hcvert_{ac}^x h_{cb}^x h_{ba}^{x}}{\ene_{ac}+2i\eta}\qty(\frac{f_{ab}}{\ene_{ab}+\Omega+i\eta} + \frac{f_{cb}}{\ene_{bc} - \Omega +i\eta} + (\Omega \to -\Omega)) \right] \label{eq:response_coeff_numerics}\\
	&= \frac{1}{V}\sum_{\vec{k}}\frac{e^2}{\Omega^2} 
	\left[\sum_{a}\hcvert_{aa}^{xxx} f_a 
	+ \sum_{a,b}\hcvert_{ab}^{xx} h_{ba}^x \qty(\frac{f_{ab}}{\ene_{ab}+\Omega} + \frac{f_{ab}}{\ene_{ab}-\Omega} - i\pi f_{ab} (\delta(\ene_{ab}-\Omega)+\delta(\ene_{ab}+\Omega)))  
	+ \sum_{a,b}\frac{\hcvert_{ab}^{x} h_{ba}^{xx}}{\ene_{ab}}f_{ab} \right. \nonumber \\ 
	& \left. + \sum_{a,b,c}\frac{\hcvert_{ac}^x h_{cb}^x h_{ba}^{x} }{\ene_{ac}}\qty(\frac{f_{ab}}{\ene_{ab}+\Omega} + \frac{f_{cb}}{\ene_{bc} - \Omega}) 
	- i\pi \sum_{a,b,c} \frac{\hcvert_{ac}^x h_{cb}^x h_{ba}^{x}}{\ene_{ac}+2i\eta} \qty(f_{ab}\delta(\ene_{ab}+\Omega) + f_{cb} \delta(\ene_{bc}-\Omega) + (\Omega \to -\Omega)) \right], \label{eq:response_coeff_general}
\end{align}
\end{widetext}
where $\ene_a=\ene_{\vec{k},a}$ is the band dispersion, $f_a=f(\ene_a), f(\ene) = (\exp(\beta(\ene-\mu))+1)^{-1}, f_{ab}=f_a-f_b, \ene_{ab} = \ene_a-\ene_b$ and $T, \mu$ are the temperature and the chemical potential. In the last line, divergences due to terms such as $1/\ene_{ab}$ with $\ene_a = \ene_b$ should be interpreted as zero. This is a general expression for shift heat current in noninteracing electronic systems.
By replacing $\hcvert$ with $h$, we can see that the expression for $\res^{xxx}$ reduces to the expression for the electric current response derived in Ref.~\cite{Parker2019}.
The last term in Eq.~\eqref{eq:response_coeff_general} diverges as $\propto 1/\eta$ if TRS is broken. This term can be interpreted as the heat current analog of the injection current~\cite{sipe2000, Ahn2020, Parker2019}, and should explicitly depend on the scattering rate. 
We also note that the heat current observed experimentally should include other contributions. For example, impurities and disorder can affect the momentum distribution of electrons resulting in finite ballistic contribution to the dc heat current as in the case of the electric current~\cite{Belinicher1980, Belinicher1982, Sturman2020, Golub2020}. Some of these contribution does not necessarily depend on the scattering rate explicitly~\cite{Belinicher1980, Golub2020} as in the case of the side-jump contribution to the anomalous Hall effect~\cite{Nagaosa2010}. Therefore, it would be difficult to fully distinguish the shift heat current contribution and other contribution due to the scattering in experiments.

\subsection{Shift heat current under TRS}
If the system has TRS, i.e., the Hamiltonian satisfies 
\begin{align}
	H_0(\vec{k})^T = H_0(-\vec{k}),
\end{align}
then the matrix element $\hcvert$ and $h$ have the following symmetry:
\begin{align}
	&(\hcvert_{\vec{k}}^{(2n+1)})^T = -\hcvert_{-\vec{k}}^{(2n+1)}, \\
	&(\hcvert_{\vec{k}}^{(2n)})^T = \hcvert_{-\vec{k}}^{(2n)}, \\
	&(h_{\vec{k}}^{(2n+1)})^T = -h_{-\vec{k}}^{(2n+1)}, \\
	&(h_{\vec{k}}^{(2n)})^T = h_{-\vec{k}}^{(2n)}.
\end{align}
With these symmetry properties, most of the terms in Eq.~\eqref{eq:response_coeff_general} cancel and only the terms with a delta-function remain. Specifically, $\res^{xxx}(2i\eta; \Omega+i\eta, -\Omega+i\eta)$ will be 
\begin{align}
	&\res^{xxx}(2i\eta; \Omega+i\eta, -\Omega+i\eta) \nonumber \\
	&= \frac{1}{V}\sum_{\vec{k}}\frac{- i\pi e^2}{\Omega^2} 
	\left[ \sum_{a,b}\hcvert_{ab}^{xx} h_{ba}^x f_{ab} (\delta(\ene_{ab}-\Omega) + \delta(\ene_{ab}+\Omega)) \right. \nonumber \\
	& + \sum_{a,b,c}\hcvert_{ac}^x h_{cb}^x h_{ba}^{x} \frac{1}{\ene_{ac}+2i\eta} \left(f_{ab}\delta(\ene_{ab}+\Omega) + f_{cb} \delta(\ene_{bc}-\Omega) \right. \nonumber \\
	& \left. + (\Omega \to -\Omega)) \right].
\end{align}
Noting that the last term vanishes due to TRS when $a=c$, one can remove the $2i\eta$ in the denominator of the last term. By using TRS, this can be rewritten as  
\begin{widetext}
\begin{align}
	&\res^{xxx}(2i\eta; \Omega+i\eta, -\Omega+i\eta) \nonumber \\
	&= \frac{1}{V}\sum_{\vec{k}}\frac{\pi e^2}{\Omega^2} 
	\left[\sum_{a,b}  \Im[\hcvert_{ab}^{xx} h_{ba}^x] f_{ab} (\delta(\ene_{ab}-\Omega)+\delta(\ene_{ab}+\Omega)) +  \sum_{a,b,c}\frac{\Im[\hcvert_{ac}^x h_{cb}^x h_{ba}^{x}]}{\ene_{ac}} \qty(f_{ab}\delta(\ene_{ab}+\Omega) + f_{cb} \delta(\ene_{bc}-\Omega) + (\Omega \to -\Omega)) \right] \nonumber \\
	&= \frac{1}{V}\sum_{\vec{k}}\frac{2\pi e^2}{\Omega^2} 
	\left[ \sum_{a,b} \Im[\hcvert_{ab}^{xx} h_{ba}^x] f_{ab} \delta(\ene_{ab}-\Omega)
	+ \sum_{a,b,c} \Im[\hcvert_{ac}^x h_{cb}^x h_{ba}^{x}] \frac{f_{ab}}{\ene_{ac}} \qty(\delta(\ene_{ab}+\Omega) + \delta(\ene_{ab}-\Omega)) \right].  \label{eq:response_coeff_TRS}
\end{align}
\end{widetext}

One can also verify that the last term for $a=b$ or $b=c$ vanishes in the presence of TRS. It is clear that for $a=b$ the term vanishes due to $f_{ab}$. For $b=c$, using the relation between $\hcvert^x$ and $h^x$,
\begin{align}
	\hcvert^x_{ab} = \frac{\ene_a+\ene_b}{2}h_{ab}^x,
\end{align}
one can conclude $\Im[\hcvert_{ab}^x h_{bb}^x h_{ba}^x]\propto\Im[\abs{h_{ab}^x}^2 h_{bb}^x] = 0$, and thus the $b=c$ terms vanish.
Therefore, the last term in Eq.~\eqref{eq:response_coeff_TRS} represents three band contributions. In contrast, the first term corresponds to two band contributions.

\subsection{Two band systems with TRS} \label{sec:numerical_results}
Let us consider a two band model with TRS. Such description is justified when a TR symmetric system is effectively described by only two bands near the Fermi level. In this case, the three band contribution can be neglected, and we obtain
\begin{align}
	& \res^{xxx}(2i\eta; \Omega+i\eta, -\Omega+i\eta) \nonumber\\
	&= \frac{1}{V}\sum_{\vec{k}}\sum_{a,b}\frac{2\pi e^2}{\Omega^2} \Im[\hcvert_{ab}^{xx} h_{ba}^x] f_{ab} \delta(\ene_{ab}-\Omega).
\end{align}
By using the Leibniz rule for the covariant derivative (Eq.~\eqref{eq:Leibniz_rule}), one can verify
\begin{align}
	&\hcvert_{ab}^{xx} = \frac{1}{2}\qty(\covd^x\covd^x[(\Emat_{\vec{k}}-\mu)^2])_{ab} \nonumber \\
	&= \qty(\frac{\ene_a+\ene_b}{2}-\mu) h_{ab}^{xx} + (h^{x}h^{x})_{ab}.
\end{align}
Because we are considering a two band case, for $a\neq b$ $\Im[\hcvert_{ab}^{xx} h_{ba}^x]$ can be rewritten as  
\begin{align}
	& \Im[\hcvert_{ab}^{xx} h_{ba}^x] = \qty(\frac{\ene_a+\ene_b}{2}-\mu) \Im[h_{ab}^{xx}h_{ba}^x] + \sum_{c}\Im[h_{ac}^{x} h_{cb}^{x}h^{x}_{ba}] \nonumber \\
	&= \qty(\frac{\ene_a+\ene_b}{2}-\mu) \Im[h_{ab}^{xx}h_{ba}^x] + \Im[h_{aa}^{x} h_{ab}^{x}h^{x}_{ba}] + \Im[h_{ab}^{x} h_{bb}^{x}h^{x}_{ba}] \nonumber \\
	&=\qty(\frac{\ene_a+\ene_b}{2}-\mu) \Im[h_{ab}^{xx}h_{ba}^x] + \Im[(h_{aa}^{x}+h_{bb}^{x}) \abs{h_{ab}^{x}}^2] \nonumber \\
	&= \qty(\frac{\ene_a+\ene_b}{2}-\mu) \Im[h_{ab}^{xx}h_{ba}^x].
\end{align}
Therefore, $\res^{xxx}(2i\eta; \Omega+i\eta, -\Omega+i\eta)$ is given by 
\begin{align}
	& \res^{xxx}(2i\eta; \Omega+i\eta, -\Omega+i\eta) \nonumber\\
	&= \frac{1}{V}\sum_{\vec{k}}\sum_{a,b}\frac{2\pi e^2}{\Omega^2} \qty(\frac{\ene_a+\ene_b}{2}-\mu) \Im[h_{ab}^{xx}h_{ba}^x] f_{ab} \delta(\ene_{ab}-\Omega).
\end{align}
Furthermore, the matrix element $\Im[h_{ab}^{xx}h_{ba}^x]$ in a two band system can be transformed into \cite{Morimoto2016, Parker2019}
\begin{align}
	& \Im[h_{ab}^{xx}h_{ba}^x] = \Im\qty[\qty(\partial_{k_x} h_{ab}^x - i\commutator{\bc^x}{h^x}_{ab})h_{ba}^x] \nonumber \\
	&= \Im\qty[\abs{h_{ab}^x}^2 \qty(\partial_{k_x}\log{h_{ab}^x} - i(\bc_{aa}^x - \bc_{bb}^x)) + i\bc_{ab} h_{ba}^x (h_{aa}^x - h_{bb}^x)] \nonumber \\
	&= \Im\qty[\abs{h_{ab}^x}^2 \qty(\partial_{k_x}\log{h_{ab}^x} - i(\bc_{aa}^x - \bc_{bb}^x)) +  \frac{\abs{h_{ab}^x}^2}{\ene_{ab}} (h_{aa}^x - h_{bb}^x)] \nonumber \\
	&= \abs{h_{ab}^x}^2 \qty(\partial_{k_x}\phi_{ab}^x - (\bc_{aa}^x - \bc_{bb}^x)) \nonumber \\
	&= \abs{h_{ab}^x}^2 R_{ab}^x,
\end{align}
where $\phi_{ab}^x=\Im\log h_{ab}^x$, and $R_{ab}^x = \partial_{k_x}\phi_{ab}^x - (\bc_{aa}^x - \bc_{bb}^x)$ is the quantity called shift vector, which also appears in the expression of the shift current~\cite{sipe2000}.  Therefore, the expression for the response coefficient is simplified to the following form:
\begin{align}
	& \res^{xxx}(2i\eta; \Omega+i\eta, -\Omega+i\eta) \nonumber \\
	&= \frac{1}{V}\sum_{\vec{k}}\sum_{a,b}\frac{2\pi e^2}{\Omega^2}  \qty(\frac{\ene_a+\ene_b}{2}-\mu) \abs{h_{ab}^x}^2 R_{ab}^x f_{ab} \delta(\ene_{ab}-\Omega). \label{eq:heat_shift_current_2bands}
\end{align}
This is almost the same form as the one for the shift current, $\sigma^{xxx}(2i\eta; \Omega +i\eta, -\Omega+i\eta)$:
\begin{align}
	&\sigma^{xxx}(2i\eta; \Omega +i\eta, -\Omega+i\eta) \nonumber \\
	&= \frac{1}{V}\sum_{\vec{k}}\sum_{a,b}\frac{2\pi e^3}{\Omega^2} \abs{h_{ab}^x}^2 R_{ab}^x f_{ab} \delta(\ene_{ab}-\Omega). \label{eq:shift_current_2bands}
\end{align}
They only differ by the factor $(\ene_a+\ene_b)/2-\mu$ in $\res$. From these expressions, one can interpret $\res^{xxx}(2i\eta; \Omega+i\eta, -\Omega+i\eta)$ as the dc heat current that originates from the difference in the intracell coordinate between conduction and valence bands. Since this coordinate difference is represented by the shift vector $R_{ab}^x$, we call this nonlinear thermal response as "shift heat current". The shift heat current does not explicitly depend on the scattering rate as in the case of the expression for the shift current Eq.~\eqref{eq:shift_current_2bands}. The independence of the scattering rate of the shift heat current might be useful to distinguish from other contributions that explicitly depend on the scattering rate. However, we again note that some contributions due to impurities and/or disorder do not necessarily depend on the scattering rate~\cite{Belinicher1980, Golub2020} and thus it would be difficult to fully distinguish the shift heat current contribution from those due to the scattering in experiments.

\subsection{Shift heat current in Rice-Mele model} \label{sec:Rice-Mele}

\begin{figure}[htbp]
    \centering
    \includegraphics[width=0.6\columnwidth]{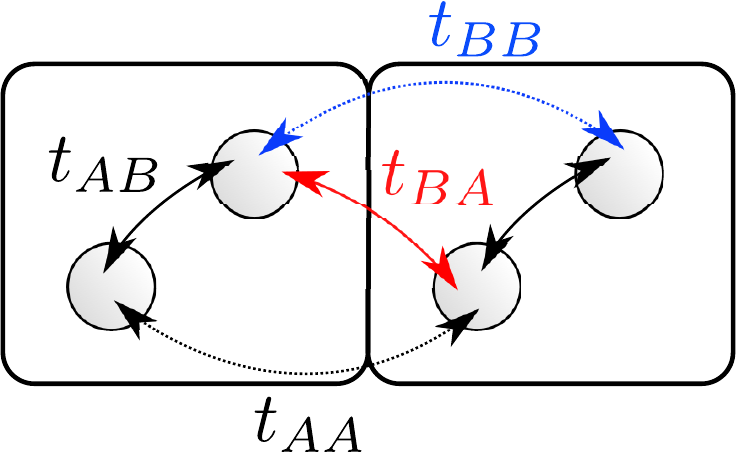}
    \caption{An illustration of Rice-Mele model (Eq.~\eqref{eq:Rice-Mele}). In order to break the particle-hole symmetry, next-nearest-neighbor hoppings $t_{AA}, t_{BB}$ are included in the model.}
    \label{fig:Rice-Mele_Model}
\end{figure}

In this section, to exemplify our theory, we present a numerical calculation of the shift heat current for Rice-Mele model~\cite{Rice1982} which is a representative 1D model with broken inversion symmetry. The Rice-Mele model can be used to describe, for example, one dimensional dimerized systems~\cite{Rice1982, Onoda2004}, single-layer monochalcogenides~\cite{Rangel2017}, and Perovskite materials \cite{Egami1993}.
The Hamiltonian for the Rice-Mele model is given by
\begin{align}
	\hat{H} &= \sum_{n} \qty(t_{AB} \cre{c}_{nB} \cre{c}_{nA} + t_{BA}\cre{c}_{n+1, A}\ani{c}_{nB} + h.c.) \nonumber \\
	&+ \sum_{n} \qty(t_{AA} \cre{c}_{n+1, A}\ani{c}_{nA} + t_{BB} \cre{c}_{n+1, B}\ani{c}_{n, B} + h.c.) \nonumber \\
	&+ \sum_{n} \qty[\qty(\ene_0+\frac{\Delta}{2})\cre{c}_{nA}\ani{c}_{nA} + \qty(\ene_0-\frac{\Delta}{2}) \cre{c}_{nB}\ani{c}_{nB})]. \label{eq:Rice-Mele}
\end{align}  
Here $n$ is the index for unit cells, $A$ and $B$ represent the two sites in a unit cell, and their positions in the unit cell are given by $r_A, r_B$. The system breaks the inversion symmetry when, for example, $\Delta\neq 0$ and $t_{AB}\neq t_{BA}$ and thus nonvanishing shift current and shift heat current appear. The second-nearest-neighbor hopping is necessary to break particle-hole symmetry; otherwise heat current becomes zero (see Eq.~\eqref{eq:heat_shift_current_2bands}).   We show a schematic picture of Rice-Mele model in Fig.~\ref{fig:Rice-Mele_Model}.

We perform the numerical calculation of $\res^{xxx}(2i\eta; \Omega+i\eta, -\Omega+i\eta)$ by directly applying the expression Eq.~\eqref{eq:response_coeff_numerics}. The energy broadening $\eta$ needs to be large enough compared to the spacing between adjacent energy levels $\Delta \ene$ but small enough compared to the other energy scale. $\Delta \ene$ can be estimated as $\Delta \ene \sim t/N$ with the energy scale of the system $t$ and the number of unit cell $N$, and thus the condition for $\eta$ is $t/N \ll \eta \ll t$. We also perform the calculation for $\sigma^{xxx}(2i\eta; \Omega+i\eta, -\Omega+i\eta)$ in the same manner. 

\begin{figure*}[htbp]
	\centering
	\includegraphics[width=2\columnwidth]{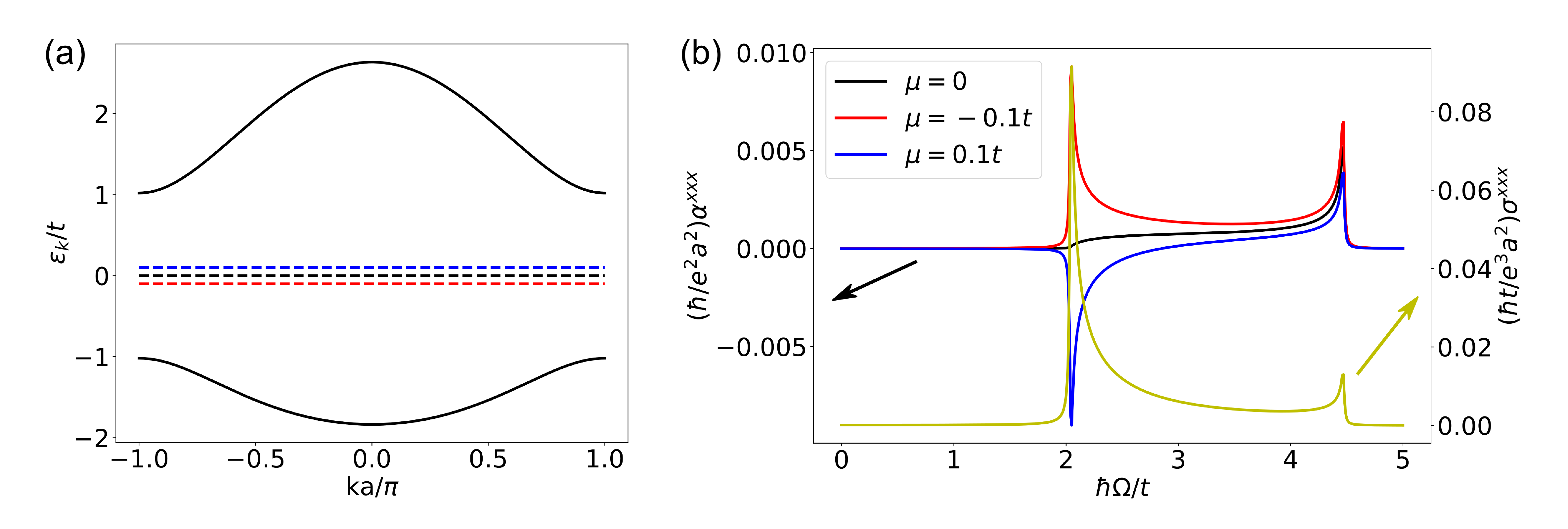}
	\caption{The shift heat current and the shift current for the Rice-Mele model. We used the parameters: $t_{AB}=0.9t, t_{BA}=1.1t, t_{AA}=0.1t, t_{BB}=0.1t, r_{A}=0, r_{B}=0.5a, \Delta=2t, \ene_0 = -0.20$ with lattice constant $a$ and an energy scale $t$. $\ene_0$ is determined so that the middle of the minimum of the conduction band and the maximum of the valence band is zero. The temperature $T$ is set to be zero ($T=0$). The total number of the unit cells $N$ is set to be 10001, and we set $\eta=0.01t$ so that $t/N \ll \eta \ll t$ with the energy scale $t$. (a) The band structure of the Rice-Mele model for the given parameters. The black, red and blue dashed lines correspond to the chemical potential $\mu=0, -0.1t, 0.1t$ respectively. (b) The response coefficients for shift heat current $\res^{xxx}(2i\eta; \Omega+i\eta, -\Omega+i\eta)$ and shift current $\sigma^{xxx}(2i\eta; \Omega+i\eta, -\Omega+i\eta)$ as a function of the photon energy $\hbar\Omega$. $\res^{xxx}$ is calculated for $\mu=0,-0.1t,0.1t$ and $\sigma^{xxx}$ is calculated for $\mu=0$.
	}
	\label{fig:numerical_results}
\end{figure*}

In Fig.~\ref{fig:numerical_results}, we show the band structure, the response coefficients $\res^{xxx}(2i\eta; \Omega+i\eta, -\Omega+i\eta)$ and $\sigma^{xxx}(2i\eta; \Omega+i\eta, -\Omega+i\eta)$ for a parameter set and several chemical potentials $\mu$ shown in the caption of the Fig.~\ref{fig:numerical_results}. We set the second-nearest neighbor hopping $t_{AA}=t_{BB}=0.1t$, and the other parameters are set following Ref.~\cite{Egami1993} where ferroelectric Perovskite BaTiO$_3$ is discussed with the Rice-Mele model (without second-nearest-neighbor hopping) and the Hubbard interaction. One can see from Fig.~\ref{fig:numerical_results} that both the shift current and the shift heat current becomes large at the band gap around $\hbar\Omega\sim t$ as expected from Eq.~\eqref{eq:heat_shift_current_2bands} and Eq.~\eqref{eq:shift_current_2bands}. When $\mu$ is set to zero, i.e., the middle of the conduction band and the valence band, the shift heat current is small reflecting the factor of $(\ene_a+\ene_b)/2-\mu$ in Eq.~\eqref{eq:response_coeff_TRS}. By varying $\mu$, the shift heat current changes even its sign. We discuss on the sign and the chemical potential dependence of the shift heat current in Sec.~\ref{subsec:sign} and Sec.~\ref{subsec:chemical_potential}.

\section{Phonon-induced shift heat current} \label{sec:phonon_induced_shift_heat_current}
Based on the diagrammatic method that we established for calculation of general heat current responses, in this section, we apply the formalism to electron-phonon coupled systems. Recently, it is experimentally shown that the optical excitation of phonons in a semiconductor induces large shift current responses in THz regions \cite{Okamura2022}. Specifically, a semiconductor BaTiO$_3$ shows large second order responses to an electric field with frequency much smaller than its band gap energy. This can be understood as a phonon-induced shift current \cite{Okamura2022}. 

The physical picture for the phonon-induced shift current is as follows. In general, excitation of phonons in noncentrosymmetric systems is accompanied by finite electronic polarization $P$ due to the electron-phonon coupling. When phonons are excited by illumination of light, the number of phonons will increase in time, and thus the polarization $P$ also increases accordingly even at steady state in noncentrosymmetric systems. Since the polarization $P$ is related to an electric current $J$ through $J=dP/dt$, the excitation of phonons results in a dc electric current. A more detailed theoretical description is given in Ref.~\cite{Okamura2022}. It should be noted that the electrons are excited only virtually in this mechanism. This is in sharp contrast to, for example, a proposal in Ref.~\cite{Budkin2020}, where phonons create real excitations of electrons in a narrow gap quantum well. If the band gap of the system is sufficiently large compared to the energy of phonons, the phonon cannot create real excitations of electrons, in which case the contributions studied in Ref.~\cite{Okamura2022} and the present paper are dominant.

From the analogy between shift current and shift heat current, we can also expect that the shift heat current can be induced by phonon excitations. Namely, we expect "electronic heat polarization" $P_Q$ is induced through electron-phonon coupling along with $P$ when phonons are excited, and it also increases in time, resulting in finite dc heat current $J_Q=dP_Q/dt$. 
\begin{figure}[htbp]
	\centering
    \includegraphics[width=\columnwidth]{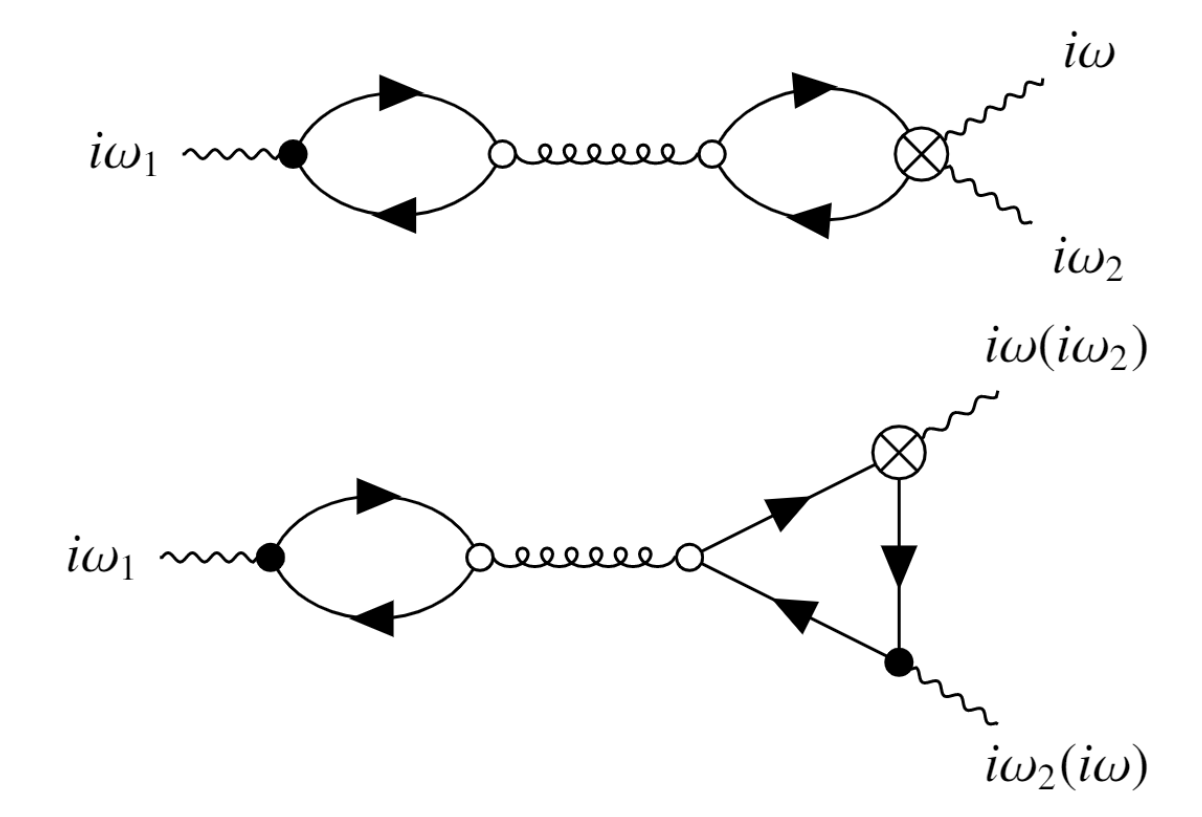}
	\caption{The Feynman diagrams which describe the phonon-induced shift heat current. Here, curly lines are phonon propagators and open dots are the electron-phonon interaction. 
	Solid lines represent propagators of electrons. Black dots and crossdots represent vertices for $h$ and $\hcvert$ as in Fig.~\ref{fig:diagram_shift_heat_current}.}
	\label{fig:phonon-induced_heat_shift_current}
\end{figure}
Following Ref.~\cite{Okamura2022}, let us calculate the phonon-induced shift heat current. We consider only one mode of phonon with wavevector $\vec{q}=0$ for simplicity. The Hamiltonian for the phonon is 
\begin{align}
	\hat{H}_{\mathrm{ph}} &= \eneph \cre{a}\ani{a},
\end{align}
where $\ani{a}$ ($\cre{a}$) is the annihilation (creation) operator of the phonon, and $\eneph$ is the energy of the phonon.
The electron-phonon interaction is described by the following Hamiltonian:
\begin{align}
	\hat{H}_{\mathrm{el-ph}} &= \sum_{\vec{k}} \cre{c}_{\vec{k}} \frac{\lambda_{\vec{k}}}{\sqrt{V}} \ani{c}_{\vec{k}} (\cre{a} + \ani{a}),
\end{align}
where $\lambda_{\vec{k}}$ is an $s\times s$ hermitian matrix ($\vec{k}$ in $\lambda_{\vec{k}}$ is often suppressed below). The diagrams corresponding to phonon-induced shift heat current are shown in Fig.~\ref{fig:phonon-induced_heat_shift_current}. By calculating these diagrams, $\res^{xxx}_{\mathrm{ph}}(i\omega; i\omega_1, i\omega_2)$ is obtained as
\begin{align}
	\res^{xxx}_{\mathrm{ph}}(i\omega; i\omega_1, i\omega_2) &= A(i\omega_1) D(i\omega_1) B_1(i\omega_1, i\omega_2) 
	\nonumber \\
	&+ A(i\omega_1) D(i\omega_1) B_2(i\omega_1, i\omega_2) 
	\nonumber \\
	&+ (i\omega_1 \leftrightarrow i\omega_2),
\end{align}
with
\begin{align}
	A(i\omega_1) &= \frac{1}{V}\sum_{\vec{k}}\sum_{a,b,\ene_n} \frac{ie}{i\omega_1} h_{ab}^x G_a(i\ene_n + i\omega_1) G_b(i\ene_n) \lambda_{ba} 
	\nonumber\\
	&= \frac{1}{V}\sum_{\vec{k}}\sum_{a,b}\frac{ie}{i\omega_1} \lambda_{ba} h_{ab}^x  I_2^{ba}(i\omega_1), \\
	B_1(i\omega_1, i\omega_2) & = \frac{1}{V}\sum_{\vec{k}}\sum_{a,b,\ene_n} \lambda_{ab} G_a(i\ene_n+i\omega_1) \frac{ie}{i\omega_2}\hcvert_{ba}^{xx} G_b(i\ene_n) 
	\nonumber\\
	&= \frac{1}{V}\sum_{\vec{k}}\sum_{a,b,\ene_n} \frac{ie}{i\omega_2}\lambda_{ab} \hcvert_{ba}^{xx} I_2^{ba}(i\omega_1),
\end{align}
\begin{align}
 	&B_2(i\omega_1, i\omega_2) \nonumber \\
 	&= \frac{1}{V}\sum_{\vec{k}}\sum_{a,b,c,\ene_n} \frac{ie}{i\omega_2} G_a(i\ene_n) \hcvert^x_{ac} G_c(i\ene_n + i\omega) \lambda_{cb} G_b(i\ene_n + i\omega_2) h_{ba}^x  
    \nonumber \\
 	&+ \frac{1}{V}\sum_{\vec{k}}\sum_{a,b,c,\ene_n} \frac{ie}{i\omega_2} G_a(i\ene_n) \hcvert^x_{ca} G_c(i\ene_n - i\omega) \lambda_{bc} G_b(i\ene_n - i\omega_2) h_{ab}^x \nonumber \\
	& = \frac{1}{V}\sum_{\vec{k}}\sum_{a,b,c} \frac{ie}{i\omega_2} [\hcvert^x_{ac}\lambda_{cb} h_{ba}^x I_3^{abc}(i\omega_2, i\omega_1) 
	\nonumber \\
 	&\hspace{6em}
 	+ \hcvert^x_{ca}\lambda_{bc} h_{ab}^x I_3^{abc}(-i\omega_2, -i\omega_1)],
\end{align}
and $D(i\omega)$ is the propagator of the phonon, 
\begin{align}
	D(i\omega) &= \frac{1}{i\omega - \eneph} - \frac{1}{i\omega + \eneph}.
\end{align}
By analytic continuation $i\omega_1\to \Omega + i\eta, i\omega_2 \to -\Omega + i\eta, i\omega \to 2i\eta$, we obtain the phonon-induced shift heat current. In the following, we assume that $\Omega(>0)$ is much smaller than the band gap and of the order of the phonon energy, $\eneph$.

In the presence of TRS, $\lambda_{\vec{k}}$ satisfies
\begin{align}
	\lambda_{\vec{k}}^T = \lambda_{-\vec{k}}.
\end{align} 
After the analytic continuation, the following relations hold:
\begin{align}
    A(\Omega) &= A(\Omega+i\eta) = A(-\Omega+i\eta) = A(-\Omega),\\
    B_i(\Omega,-\Omega) &= B_i(\Omega+i\eta, -\Omega+i\eta) = -B_i(-\Omega+i\eta, \Omega+i\eta) \nonumber\\
    &= -B_i(-\Omega, \Omega) \quad (i=1,2),
\end{align}
for inifinitesimal $\eta$.
Therefore, the expression for $\res_{\mathrm{ph}}^{xxx}$ after the analytic continuation reduces to
\begin{align}
    &\res^{xxx}_{\mathrm{ph}}(2i\eta; \Omega+i\eta, -\Omega+i\eta) \nonumber\\
    &= -2\pi i\delta(\Omega-\eneph) A(\Omega)\qty(B_1(\Omega, -\Omega) + B_2(\Omega, -\Omega)). \label{eq:phonon-induced_heat_shift_current}
\end{align}

For simplicity, we consider a two-band case with TRS below. By straightforward calculation, we obtain 
\begin{align}
    A(\Omega) &= \frac{e}{V}\sum_{\vec{k}}\sum_{a,b} \Im[\lambda_{ba}h_{ab}^x]\frac{f_{ba}}{\ene_{ba}^2-\Omega^2}, \\
    B_1(\Omega) &= \frac{-ie}{V\Omega}\sum_{\vec{k}} \sum_{a,b} \Re[\lambda_{ab} \hcvert_{ba}^{xx}]\frac{f_{ba}\ene_{ba}}{\ene_{ba}^2-\Omega^2},\\
    B_2(\Omega) &= \frac{2ie}{V\Omega}\sum_{\vec{k}}\sum_{a<b} f_{ab}
	\left[\frac{1}{\ene_{ab}^2 - \Omega^2} \Re[\hcvert^x_{ab}\lambda_{ba}] (h_{aa}^x-h_{bb}^x) \nonumber \right. \\ 
	& + \frac{1}{\ene_{ab}^2 - \Omega^2} \Re[\hcvert^x_{ba} h_{ab}^x] (\lambda_{aa}- \lambda_{bb}) \nonumber\\
	& \left. + \frac{\ene_{ab}^2+\Omega^2}{(\ene_{ab}^2-\Omega^2)^2} \Re[\lambda_{ab} h_{ba}^x](\hcvert^x_{aa}-\hcvert^x_{bb}) \right].
\end{align}

This is almost of the same form as the phonon-induced shift current. One can obtain the expression for the phonon-induced shift current by replacing $\hcvert$ by $h$. 

\section{Discussions} \label{sec:discussion}
\subsection{Symmetry condition to observe shift heat current}
In order to observe shift heat current, the system needs to break the spatial inversion symmetry. This is obvious from symmetry consideration of the nonlinear response tensor, but we can also confirm explicitly that the response coefficient $\res^{xxx}$ vanishes in centrosymmetric systems as follows. If the system preserves the spatial inversion symmetry, the Hamiltonian satisfies
\begin{align}
	H_0(-\vec{k}) &= P^\dagger H_0(\vec{k}) P,
\end{align}
where $P$ is the unitary matrix which expresses the spatial inversion. For example, in the case of Rice-Mele model in Sec.~\ref{sec:numerical_results}, $P=\sigma_x$ where $\sigma_x$ is the Pauli matrix. 

Under the inversion symmetry, the matrix element $\hcvert$ and $h$ have the following symmetry:
\begin{align}
	&\hcvert_{\vec{k}}^{(2n+1)} = -\hcvert_{-\vec{k}}^{(2n+1)}, \\
	&\hcvert_{\vec{k}}^{(2n)} = \hcvert_{-\vec{k}}^{(2n)}, \\
	&h_{\vec{k}}^{(2n+1)} = -h_{-\vec{k}}^{(2n+1)}, \\
	&h_{\vec{k}}^{(2n)} = h_{-\vec{k}}^{(2n)},
\end{align}
where $\hcvert_{\vec{k}}^{(m)}$ denotes $(1/2)(\covd^x)^m[\Emat_{\vec{k}}^2]$ and similar for $h_{\vec{k}}^{(m)}$. We also explicitly show the $\vec{k}$-dependence of $\hcvert$ and $h$. Therefore, all the terms in Eq.~\eqref{eq:response_coeff_general} are odd in $\vec{k}$ and vanish after the $\vec{k}$-summation. This means that the spatial inversion symmetry breaking is necessary to induce the second order response of the heat current, as in the case of the electric current. 

One can also verify that the phonon-induced shift heat current vanishes under the inversion symmetry by using $\lambda_{\vec{k}} = \lambda_{-\vec{k}}$. 

\subsection{Sign of shift heat current} \label{subsec:sign}
The sign of shift heat current is determined by two factors. One is the shift vector or equivalently the polarity of the system, and the other is the factor of $(\ene_a+\ene_b)/2-\mu$ appearing in Eq.~\eqref{eq:heat_shift_current_2bands}. The shift vector represents the inversion symmetry breaking of the system, and it also determines the direction of shift current. If the polarity of the system is reversed, the shift vector will be also reversed, hence shift heat current changes its sign. In noncentrosymmetric systems, the inversion symmetry breaking often results in finite polarization. In that case, the polarity of the system and the direction of shift heat current can be reversed by applying an electric field. This property is the same as shift current.

The other factor, $(\ene_a+\ene_b)/2-\mu$, is characteristic to heat transport and its sign is determined in a similar way to the Seebeck coefficient. Let us consider the case of insulators with finite band gap. If $(\ene_a+\ene_b)/2-\mu < 0$ where $\ene_a$ represents a conduction band while $\ene_b$ represents a valence band, the chemical potential is closer to the conduction band. Therefore, the situation is quite similar to a Seebeck effect where the dominant carrier is electron and the Seebeck coefficient is negative. However, we note that we are considering insulators here and thus there is almost no carriers nor the concept of the dominant carrier itself. A more appropriate interpretation may be the following: in the shift heat current, a pair of an electron and a hole excited by an incident photon carries the heat of $(\ene_a+\ene_b)/2-\mu$ in total. In this case, the number of excited electrons and holes are the same, but the net heat carried by them is finite. 

\subsection{Chemical potential and temperature dependence of shift heat current} \label{subsec:chemical_potential}
As seen from the definition of the heat current Eq.~\eqref{eq:heat_current_operator_def}, the heat current explicitly depends on the chemical potential $\mu$.
In the case of linear dc responses in metals or doped semiconductors, only the carriers in levels near $\mu$ contributes to the transport and the change of $\mu$ results in change of both the electric current and the heat current. In contrast, shift current and shift heat current occur in insulators. In this case, as long as the change of Fermi distribution function due to the change of chemical potential is negligible, the shift current is almost independent of $\mu$ while the shift heat current does depend on $\mu$, as seen from Eq.~\eqref{eq:shift_current_2bands} and Eq.~\eqref{eq:heat_shift_current_2bands}. Therefore, if the chemical potential is shifted by $\Delta\mu$ because of, for example, impurities, the shift heat current changes by $\sim(\Delta \mu/e) J_e^{(\mathrm{shift})}$ with shift current $J_e^{(\mathrm{shift})}$. (Here we neglect the change of the shift current.) In other words, one can control the shift heat current by varying the chemical potential.
Since the chemical potential in insulators strongly depends on the existence of impurities, we can also expect that a small amount of impurities induces a large shift of the chemical potential and thus drastically changes the magnitude of the shift heat current. 

The temperature dependence of the shift heat current in insulators is determined almost only by the temperature dependence of the chemical potential. If the change of temperature $\Delta T$ satisfies $\kB\Delta T \ll E_G$, then the shift heat current Eq.~\eqref{eq:heat_shift_current_2bands} changes by $\sim(\Delta \mu/e) J_e^{(\mathrm{shift})}$, where $\Delta\mu$ is the change due to the temperature variation.

For example, if the density of states of conduction band and that of valence band are respectively given by $D_c(\ene+\ene_{c0})=A_c\ene^{s_c}$ and $D_v(\ene_{v0}-\ene)=A_v\ene^{s_v}$ where $\ene_{c0(v0)}$ is the bottom (top) of the conduction (valence) band and $A_c, A_v, s_c, s_v$ are constants. If one further assumes that $\ene_{c0}-\mu \gg \kB T$, $\mu-\ene_{v0} \gg \kB T$ and the charge neutrality condition 
\begin{align}
    \int_{\ene_{c0}}^\infty f(\ene)D_c(\ene) \dd{\ene} = \int_{-\infty}^{\ene_{v0}} (1-f(\ene))D_v(\ene) \dd{\ene},
\end{align}
then the temperature dependence of chemical potential is 
\begin{align}
    \mu \simeq \frac{\ene_{v0}+\ene_{c0}}{2} + \frac{\kB T}{2}\qty[\log(\frac{A_v}{A_c} \frac{\Gamma(s_v+1)}{\Gamma(s_c+1)}) + (s_v-s_c)\log\kB T].
\end{align}
Here, $\Gamma(x)=\int_0^\infty e^{-t}t^{x-1} \dd{t}$ is the Gamma function.
In particular, if both the conduction and valence band are parabolic with effective mass $m_c$, $m_v$, then $s_c=s_v=(d-2)/2$ with the spatial dimension $d$ and 
\begin{align}
    \mu \simeq \frac{\ene_{v0}+\ene_{c0}}{2} + \frac{d}{4}\kB T\log\frac{m_v}{m_c}.
\end{align}
Therefore, $\Delta\mu \propto \Delta T$ and thus the shift heat current is linearly dependent on the temperature in this case.

\subsection{Candidate materials and order estimation}
Materials that break inversion symmetry can support nonzero shift heat current responses. Because of the similarity between shift heat current (Eq.~\eqref{eq:heat_shift_current_2bands}) and shift current (Eq.~\eqref{eq:shift_current_2bands}), one can expect that materials which exhibits large shift current responses also shows large shift heat current responses. Namely, 
\begin{align}
	J_Q^{(\mathrm{shift})} \sim \frac{t}{e} J_e^{(\mathrm{shift})}, \label{eq:order_estimation}
\end{align}
where $J_Q^{(\mathrm{shift})}$ and $J^{(\mathrm{shift})}$ are the shift heat current and the shift current, and $t$ is the characteristic energy scale corresponding to the factor $(\ene_a+\ene_b)/2-\mu$ in Eq.~\eqref{eq:heat_shift_current_2bands}. As mentioned at the end of Sec.~\ref{subsec:sign}, the factor $t$ can be interpreted as a net heat carried by the excited electron and hole, and Eq.~\eqref{eq:order_estimation} is a generalization of Eq.~\eqref{eq:J_JQ_relation} to the second order response. We can expect that the relation \eqref{eq:order_estimation} holds also for general multiband systems with an appropriate modification of an expression for $t$, because the general expression for $\alpha^{xxx}$ (Eq.~\eqref{eq:response_coeff_general}) differs from $\sigma^{xxx}$ only by $\tilde{g}$ in place of $h$.

One representative material which shows the shift current response is SbSI \cite{Ogawa2017, Sotome2019}. SbSI exhibits shift current of the order of $\SI{0.1}{\nano\ampere}$ when illuminated by a cw laser with $\hbar\omega=\SI{1.95}{\electronvolt}$, and \SI{10}{\micro\ampere} at peak when irradiated by a pulsed laser with $\hbar\omega=\SI{2.05}{\electronvolt}$ and the power $\SI{0.6}{\micro\joule}$~\cite{Ogawa2017}. In this case, as the energy scale $t$ is estimated as $t\sim\SI{1}{\electronvolt}$, the expected shift heat current is $J_Q^{(\mathrm{shift})} \sim \SI{0.1}{\nano\watt}$ for the cw excitation, while $J_Q^{(\mathrm{shift})}\sim\SI{10}{\micro\watt}$ at peak is expected when the laser pulse is applied. 

Another promising candidate is TaAs. TaAs is a Weyl semimetal with broken inversion symmetry that shows large shift current responses~\cite{Osterhoudt2019a}. It is experimentally observed that a photocurrent $\sim \SI{1}{\micro\ampere}$ can be induced by a laser with wavelength $\SI{10.6}{\micro\metre}$ (\SI{117}{\milli\electronvolt}) and power $\sim\SI{100}{mW}$. 
If we assume that the Weyl semimetal TaAs has Hamiltonian of the form $H = v_0 k +  v^{\mu\nu}k^{\mu}\sigma^{\nu}$ where $\mu, \nu$ run over $x,y,z$, then $t$ for photon with energy $\hbar\omega$ can be estimated as $t \sim (v_0/2v)\hbar\omega -\mu$. Furthermore we estimate $\frac{v}{2v_0}\sim 0.2$ and $\mu\sim\SI{10}{\milli\electronvolt}$~\cite{Lee2015a}, then $t\sim\SI{10}{\milli\electronvolt}$ and one obtains from Eq.~\eqref{eq:order_estimation},
\begin{align}
	J_Q^{(\mathrm{shift})} \sim \SI{10}{\nano\watt}.
\end{align}

As for the phonon-induced shift heat current, a material which exhibits a large shift current is promising as well. Since the expressions for the phonon-induced shift heat current is different from that of the electric current only by the factor $\hcvert$ in the place of $h$,  Eq.~\eqref{eq:order_estimation} also holds for the phonon-induced currents. We emphasize that $t$ is determined by the energy scale of electrons even for the phonon-induced currents. 
In an experiment, it is observed that BaTiO$_3$ shows a phonon-induced shift current as large as $\sim\SI{10}{\micro\ampere}$ \cite{Okamura2022}. Assuming $t\sim\SI{1}{\electronvolt}$ for BaTiO$_3$, the phonon-induced shift heat current is estimated as
\begin{align}
	J_Q^{(\mathrm{shift, ph})} \sim \SI{10}{\micro\watt},
\end{align}
which is quite large compared to the response of TaAs.

We also note that by varying chemical potential, the shift heat current in the above materials can be changed by $\sim(\Delta\mu/e) J_e^{(\mathrm{shift})}$ as discussed in Sec.~\ref{subsec:chemical_potential}.

In conclusion, we have established a diagrammatic formulation of the nonlinear heat current response to an ac electric field, and calculated the second order response, which we call shift heat current. We have derived the microscopic expression for the shift heat current and confirmed that the shift heat current is determined by the shift vector, as in the case of the shift current. The amplitude of shift heat current $J_Q^{\mathrm{(shift)}}$ is roughly estimated as $J_Q^{\mathrm{(shift)}} \sim (t/e) J_e^{\mathrm{(shift)}}$ where $t$ is the characteristic energy scale of electrons, and can be controlled by changing the chemical potential. We have also calculated the phonon-induced shift heat current and found that even for phonon-induced cases, $J_Q^{\mathrm{(shift,ph)}} \sim (t/e) J_e^{\mathrm{(shift,ph)}}$ still holds and the amplitude of $J_Q^{\mathrm{(shift)}}$ is determined by the energy scale of electrons, not that of phonons.

\acknowledgements
The authors are grateful to T. Sagawa, M. Hirschberger, and Y. Xu for useful comments and discussions. N.N. was supported by JST CREST Grant Number JPMJCR1874 and JPMJCR16F1, Japan, and JSPS KAKENHI Grant Number 18H03676. T.M. was supported by JST PRESTO (JPMJPR19L9) and JST CREST (JPMJCR19T3). Y.O. thanks the World-leading Innovative Graduate Study Program for Materials Research, Information, and Technology (MERIT-WINGS).

\bibliographystyle{apsrev4-2}
\bibliography{library}

\end{document}